\documentclass[runningheads]{llncs}

\usepackage{graphicx}
\usepackage{tabularx}
\usepackage{cite}

\usepackage{academicons}
\usepackage{hyperref}
\usepackage{xcolor}
\newcommand{\orcid}[1]{\href{https://orcid.org/#1}{\includegraphics[scale=0.08]{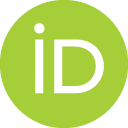}}}

\begin{document}
\title{Coding Cross Sections of an Electron Charge Transfer Process}
\titlerunning{ }
%
\author{Emília Valença Ferreira de Aragão\inst{1,2}\orcid{0000-0002-8067-0914} 
\and Luca Mancini\inst{2}\orcid{0000-0002-9754-6071}
\and Xiao He\inst{4}
\and Noelia Faginas-Lago\inst{2}\orcid{0000-0002-4056-3364} 
\and Marzio Rosi\inst{3}\orcid{0000-0002-1264-3877} 
\and Daniela Ascenzi\inst{4}\orcid{0000-0001-5393-9554}
\and  Fernando Pirani\inst{2}\orcid{0000-0003-3110-6521}
}

\authorrunning{E.V.F. Aragão et al.}
%
\institute{
Master-tec srl, Via Sicilia 41, 06128 Perugia, Italy\\
\email{emilia.dearagao@master-tec.it}\\
\and
Dipartimento di Chimica, Biologia e Biotecnologie,\\ Universit\`{a} degli Studi di Perugia, 06123 Perugia, Italy\\ 
\email{\{emilia.dearagao,luca.mancini2\}@studenti.unipg.it}\\
\email{\{noelia.faginaslago,fernando.pirani\}@unipg.it}\\
\and
 Dipartimento di Ingegneria Civile ed Ambientale,\\ Universit\`{a} degli Studi di Perugia, 06125 Perugia, Italy\\
 \email{marzio.rosi@unipg.it}\\
\and
 Dipartimento di Fisica, Università di Trento, Trento, Italy
 \email{daniela.ascenzi@unitn.it}
}

\maketitle              
\begin{abstract}
The paper presents the algorithm of a code written for computing the cross section for a charge transfer process involving a neutral molecule and a monatomic ion.
The entrance and exit potential energy surfaces, driving the collision dynamics, are computed employing the Improved Lennard-Jones function that accounts for the role of non-electrostatic forces, due to size repulsion plus dispersion and induction attraction. In addition, electrostatic components, affecting the entrance channels, are evaluated as sum of Coulomb contributions, determined by the He$^+$ ion interacting with the charge distribution on the molecular frame.
%
%
The cross section is estimated by employing the Landau-Zener-Stückelberg approach.
The code implemented has been employed in systems involving helium cation and a small organic molecule, such as methanol, dimethyl ether and methyl formate.

\end{abstract}
%
%
%

\section{Introduction}

In the last 70 years, more than 260 molecules have been detected in the interstellar medium \cite{mcguire2022census,Muller2001Cologne,MULLER2005Cologne,ENDRES2016Cologne}.
Among them, C-bearing molecules with 6 or more atoms have been given the name of interstellar Complex Organic Molecules, or simply iCOMs \cite{herbst2009complex,ceccarelli2017seeds}.
The presence of a significant number of molecules, with increasing chemical complexity (including species that can be considered precursors of more complex and prebiotic molecules, such as formamide\cite{lopez2019interstellar}) in different regions of the interstellar medium (ISM) is a challenge for our comprehension of interstellar chemistry, considering also the harsh conditions of some astronomical environments, which are considered prohibitive for molecular growth (where the temperature can go down to 10 K and the particle density can be as low as 10$^{4}$ particles cm$^{-3}$ \cite{caselli2012our}).
Several models have been implemented with the purpose of setting a complete picture for the understanding of the chemical processes leading to the synthesis of iCOMs \cite{herbst2017synthesis,agundez2013chemistry,taquet2012multilayer,garrod2006formation,vasyunin2017formation,garrod2008complex}. Currently, two different mechanisms can be invoked in order to
explain the molecular complexity of the ISM: gas phase reactions and/or grain surface processes.
The first suggested mechanisms is related to the formation of iCOMs on the surface of the grains, through hydrogenation processes, leading to the formation of simple radicals. The subsequent warming up of the surrounding environment during the first phases of the star formation, allows the radicals to acquire mobility and diffuse on the grain surface to form several iCOMs.
Grain surface chemistry cannot be invoked to reproduce the observed molecular abundances by itself. For this reason, a combined approach including gas phase reactions is implemented in the astrochemical models. Following this approach, the formation of simple hydrogenated and oxidized species (e.g. H$_2$O, NH$_3$, CO) on the surface of the grains during the first stages of star formation is followed by the ejection of the synthesized molecules in the gas phase during the hot-corino \cite{caselli2012our} stage. At this point, subsequent gas-phase reactions are the main route of formation of iCOMs. Processes might involve ions, radicals and/or closed-shell species provided that those processes are exothermic and barrierless.
Recent works \cite{balucani2015formation,skouteris2018genealogical,rosi2018possible} attest that the role of gas-phase chemistry appears indeed to be pivotal.\\
\indent
A synergistic and multidisciplinary approach is fundamental to unravel the chemistry of the ISM.
Such an approach includes astronomical observations, chemical analysis (both with experiments and theoretical quantum chemistry calculations) and astrochemical/astrophysical models, allowing a comparison between theoretical predictions and astronomical observations.
A peculiar example can be represented by two databases which are mostly used in astrochemical
models, namely KIDA\cite{Wakelam2012KIDA} and UMIST\cite{Woodall2007UMIST}, in which the abundances of detected species are accounted for by considering their formation and destruction routes.
Unfortunately, among all the processes included in the models, some processes might be missing
while many processes are included with estimated, even just guessed, parameters considered on
the basis of common sense or chemical analogies. In order to explain the abundance of a molecule, it is extremely important to properly consider, together with the formation routes, also the possible destruction reactions of iCOMs.
Among the possible reaction pathways, destruction of iCOMs can happen after a collision with atomic or molecular species.
A peculiar example is represented by the collision with high energetic ions.
When the collision happens with an ion, for instance He$^+$ or H$_3^+$ and HCO$^+$, the three most abundant ions in the ISM (generated by the interaction of H, He and H$_2$ with cosmic rays) charge or proton exchange processes might occur, followed by the dissociation of the iCOM.
The destruction of interstellar Complex Organic Molecules appears to be indeed dominated by the reactions with He$^+$, H$_3^+$ and HCO$^+$.
In particular, due to its high value of ionisation energy\cite{mallard2000nist}, helium represents a pivotal species to drive the chemistry of the ISM\cite{lepp2002atomic,de2014h+}.
The dynamics of the charge transfer process depends on the probability of the transition between different intermolecular Potential Energy Surfaces (PESs).
It is important to assess whether the collision happens with a preferential orientation, therefore a model of the cross section is proposed and their evolution is compared with experimental measurements of the cross section in order to fully understand the reaction dynamic.

In this work, we describe how the PESs have been implemented in the code written by our group, in order to calculate the cross sections for the global electron transfer process, to be compared with results measured as a function of the collision energy with the molecular beam technique, and, subsequently, to evaluate related reaction rate constants as a function of the temperature.


\section{Program Implementation}

The code has been developed to perform a theoretical analysis of the cross section for the charge exchange process, allowing to estimate the values of the rate constant at different temperatures.
All parts of the code were written in C language, and have a common structure:
\begin{enumerate}
    \item {Inclusion of useful libraries as stdio.h, math.h, stdlib.h and string.h;}
    \item {Definition of assorted functions;}
    \item {Main program section, defined inside ``\textit{int main\{...\}}".}
    
\end{enumerate} 

In the next sections, a detailed description of the contents of each piece of code is given.

\subsection{Calculating cuts of the entrance and exit potential energy surface with \textit{pes.c}}
The program has been written in order to characterize the entrance and the exit potentials (\textit{i.e.} before and after the charge transfer, respectively) of a reaction involving a neutral molecule (e.g. dimethyl ether, methyl formate, methanol) and a monoatomic ion (e.g. He$^{+}$).
The exploration of the entrance potential energy surface serves to visualize how the reactants are approaching each other: the lower the energy at the bottom of the well, the more attractive the relative spatial orientation.
Later, cuts of the entrance and exit PES are plotted to verify whether an exothermic crossing is possible.
From here on, an individual relative spatial orientation between the reactants will be referred to as a ``configuration".
\\
\indent
\begin{figure}[t]
\centering
\includegraphics[width=\linewidth]{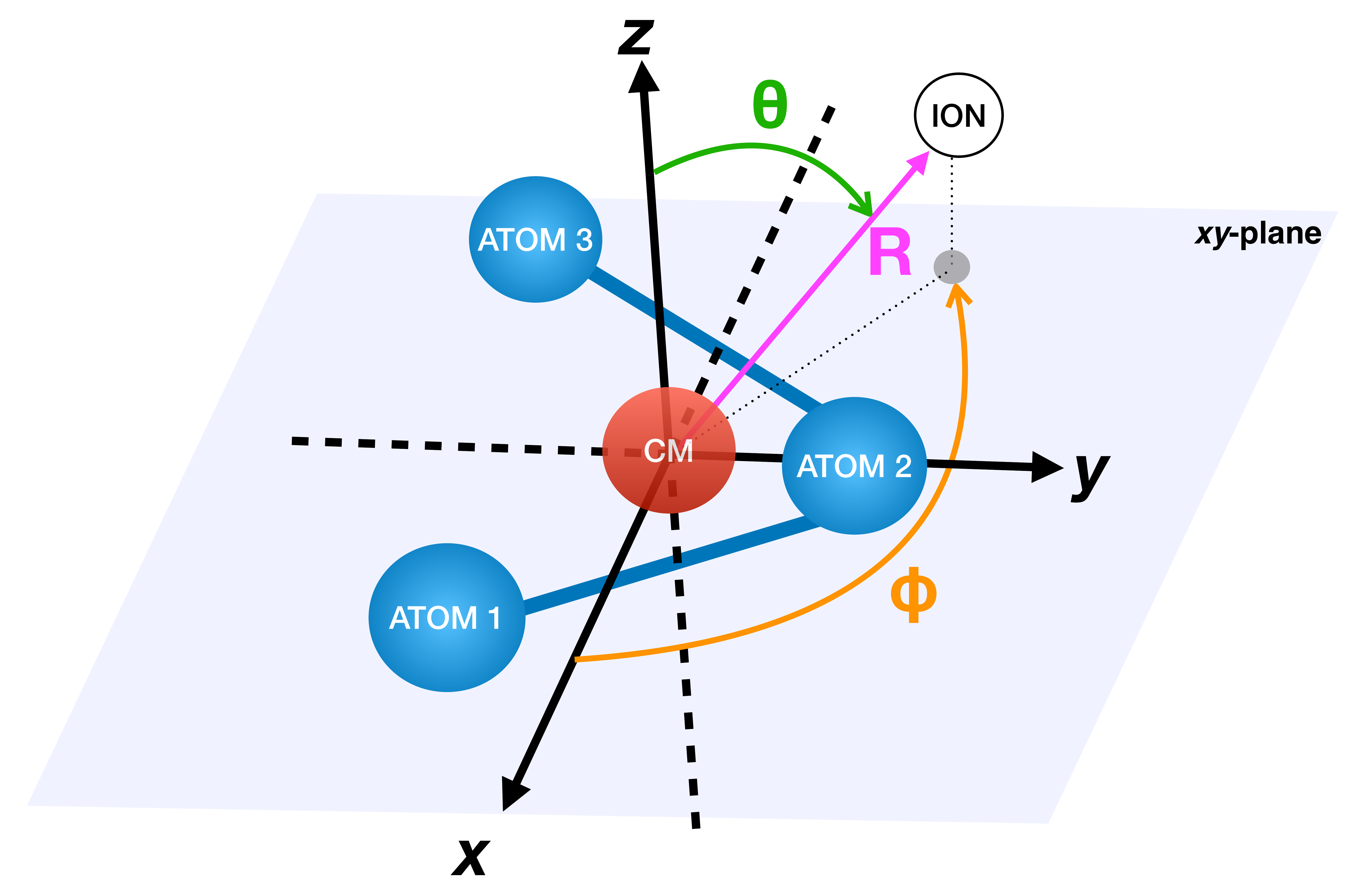}
\caption{Scheme representing a system in a three-dimensional space. Atoms in blue represent effective atoms in a neutral molecule. 
The origin of the Cartesian reference system is taken to coincide with the center of mass (CM) of the molecule. 
The placement of the ion is made by choosing the values of its spherical coordinates, from which the Cartesian coordinates can be obtained. Here, $R$ is the distance between the molecule's CM and the ion, $\theta$ is the angle made by the $R$ vector with the z-axis and $\phi$ is the azimuthal angle.}
\label {fig_space}
\end{figure}
\begin{figure}[h]
\centering
\includegraphics[width=0.95\linewidth]{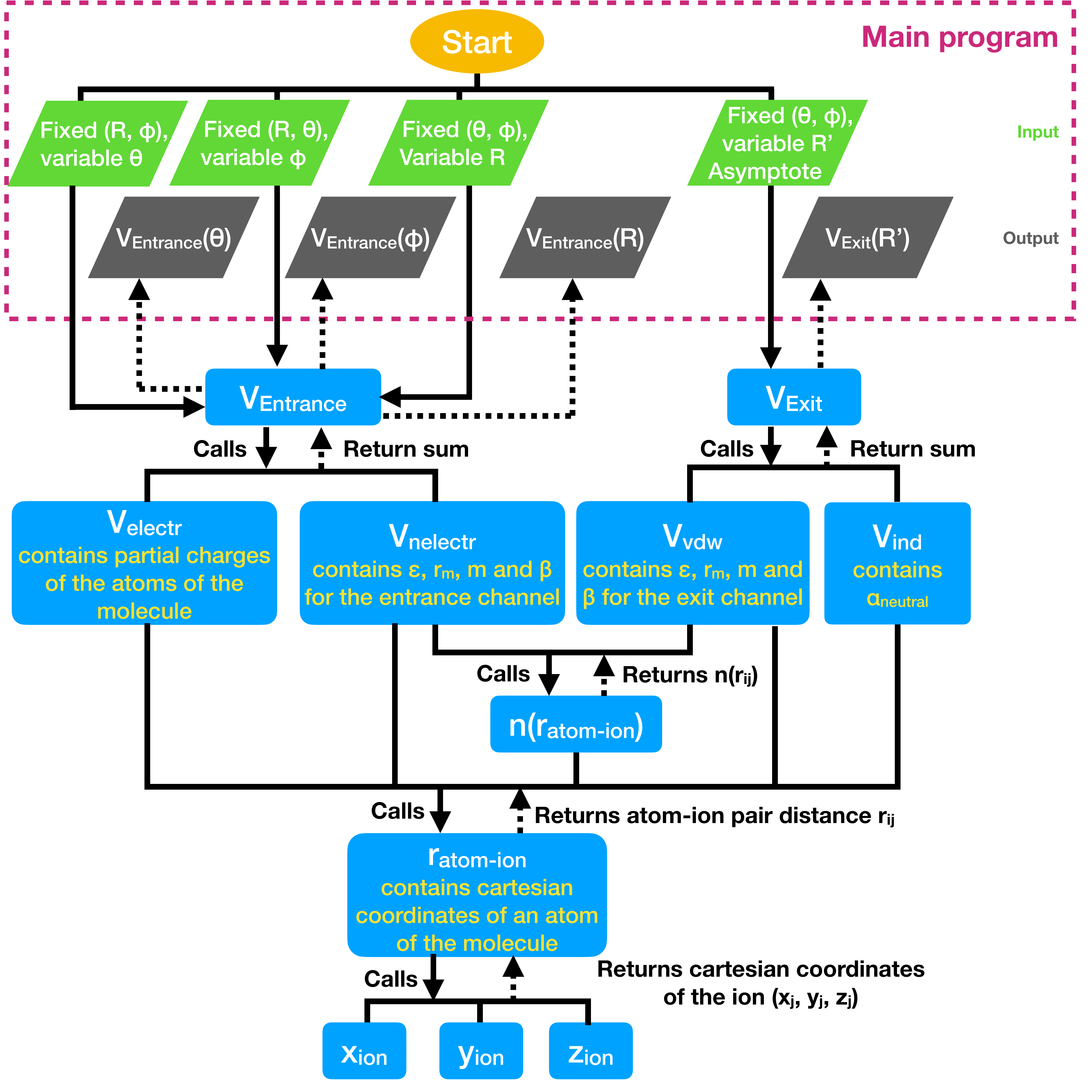}
\caption{An overall view of the algorithm in \textit{pes.c} code. In green, the input information, in grey the output. Blue blocks refer to functions that have been defined outside the main section of the program.}
\label {fig_pes}
\end{figure}
The code requires to define the position of the reactants in space.
First, the geometry of the molecule should be pre-optimized via calculations employing \textit{ab initio} methods, so intramolecular distances and bond angles are represented correctly.
Depending on the total number of atoms, it might be necessary to replace functional groups with effective atoms, for instance C$_{\mathrm{eff}}$ instead of CH$_{3}$. In this case, the partial charge of the effective atom should be the sum of the partial charges of the atoms involved.
For the whole molecule, the position of the center of mass (CM) is defined via the equations:
\begin{equation}
x_{CM} = \frac{\sum_{i=1}^{N}{x_{i}m_{i}}}{M};\;  y_{CM} = \frac{\sum_{i=1}^{N}{y_{i}m_{i}}}{M};\;  z_{CM} = \frac{\sum_{i=1}^{N}{z_{i}m_{i}}}{M}
\label{eq:1}
\end{equation}
\noindent
where $x_i$, $y_i$ and $z_i$ are the coordinates of each effective atom, $m_i$ its mass, $N$ the total number of effective atoms and $M$ the total mass of the molecule.
For an easier manipulation of the system, CM is placed at the origin of the Cartesian reference system.
The new coordinates of the effective atoms are obtained by extracting the CM coordinates from the original atomic coordinates:
\begin{equation}
x'_{i} = x_{i}-x_{CM};\; y'_{i} = y_{i}-y_{CM};\; z'_{i} = z_{i}-z_{CM}
\label{eq:2}
\end{equation}
\noindent
Once the molecular frame is fixed in the 3D space, as shown in Fig.~\ref{fig_space}, a code named ``\textit{pes.c}" can be used to set the ion position in space using spherical coordinates (see Fig.~\ref{fig_pes}). 
The Cartesian coordinates of the ion (denoted by \textit{j}) can then be derived via the coordinate transformation formulae:
\begin{equation}
x_{j} = R sin(\theta)cos(\phi)
\label{eq:3}
\end{equation}
\noindent
\begin{equation}
y_{j} = R sin(\theta)sin(\phi)
\label{eq:4}
\end{equation}
\noindent
\begin{equation}
z_{j} = R cos(\theta)
\label{eq:5}
\end{equation}
\noindent

Where $R$ is the distance between the ion and CM, $\theta$ the angle in relation with the z-axis and $\phi$ is the azimuth angle. $R$ values can range from 0 to infinite, $\theta$ values range from 0$^{\circ}$ to 180$^{\circ}$ and $\phi$ ranges from 0$^{\circ}$ to 360$^{\circ}$.
Knowing the Cartesian coordinates of the ion, the distance between each atom of the molecule and the ion can be calculated considering the following relation: 
\begin{equation}
r_{ij} = \sqrt{(x_{i}-x_{j})^2 + (y_{i}-y_{j})^2 +(z_{i}-z_{j})^2} 
\label{eq:6}
\end{equation}
\noindent

Other than spatial position of the reactants, some parameters specific to each system are necessary to be adapted before using the program.
One set of parameters are the partial charges on the neutral species that are used to estimate the electrostatic contribution for the entrance PES, $V_{electrostatic}(R)$.
Atomic charges can be obtained through \textit{ab initio} calculations, with the same level of theory used for the optimization of the molecular structure.

Other sets of parameters are adopted to define the non-electrostatic contribution $V_{ij}$, arising from each $ij$ interacting pair (again, $i$ denotes an atom of the molecule and $j$ denotes the ion) taken at a separation distance $r_{ij}$. In particular, the value of the potential well $\varepsilon_{ij}$ and of related equilibrium distance $r_{m_{ij}}$ are used to evaluate $V_{ij}$ through the Improved Lennard-Jones model \cite{pirani2008beyond}:
%
%
%
\begin{equation}V_{ij}(r_{ij}) =
\varepsilon_{ij} \left[\frac{m}{n(r_{ij})-m}\left (
\frac{r_{m_{ij}}}{r_{ij}} \right)^{n(r_{ij})} -
\frac{n(r_{ij})}{n(r_{ij})-m}\left(
\frac{r_{m_{ij}}}{r_{ij}}\right)^m \right]
\label{eq:7}
\end{equation}
\noindent
As explained in \cite{pirani04:37} and \cite{pirani2008beyond}, the $m$ parameter depends on the nature of the system: it assumes the value of 1 for ion-ion interaction, 2 for ion-permanent dipole, 4 for ion-induced dipole, and 6 for neutral-neutral pairs.
The present entrance channels involve a small ion and a neutral molecule for which the induction attraction dominates over the dispersion. Therefore, for such channels we have considered $m$ = 4. Moreover, in the exit channels (after electron transfer), a molecular ion interacts with He, a low polarizabilty atom, and the value of $m$ = 6 has been selected since the dispersion plays an important role and the small induction term is inserted as additive attractive contribution (see below).
%
%
 %
The modulation in the decline of the repulsion and the enhancement of the attraction in eq.~\ref{eq:7}, $n(r_{ij})$ takes the following form:
\begin{equation}
n(r_{ij}) = \beta + 4\left(\frac{r_{ij}}{r_{m_{ij}}}\right)^2
\label{eq:8}
\end{equation}
\noindent
where $\beta$ is a parameter related to the nature and the hardness of the interacting particles \cite{pirani04:37,bart2008,cappelletti2008,pacifici13:2668} and the $\frac{r_{ij}}{r_{m_{ij}}}$ ratio is introduced as a reduced distance. 
The Improved Lennard-Jones function is then used for both entrance and exit potentials, and in each case $\beta$ and $m$ assume values according to the nature of the system. 
The total non-electrostatic contribution, which is the sum of all $V_{ij}$ of a given configuration, is denoted $V_{nelectr}$ in the entrance channel and $V_{vdw}$ in the exit channel (see Fig.~\ref{fig_pes}).

In the entrance channel, the electrostatic component $V_{electr}(R)$, due to the interaction between the monocation He$^+$ and the anisotropic charge distribution on the molecular frame, is defined by the Coulomb's law:
\begin{equation}
V_\mathrm{electr}(R) = \frac{1}{4\pi\varepsilon_{0}} \sum_{i}^{N}\frac{q_i}{r_{ij}}
\label{eq:9}
\end{equation}
\noindent
where $R$ is again the distance between the ion and CM, $r_{ij}$ is again the distance between an atom of the molecule and the ion, obtained deconvolving $R$ into partial components, $\varepsilon_{0}$ is the vacuum permittivity, and $q_{i}$ corresponds to the partial charge on each atom of the molecule.

As indicated above, in the exit channels an induction term is inserted to represent the ion-induced dipole contribution as a pertubative approach.
The induction term describes the interaction between the newly formed cation and the neutralized species. The induction term $V_{ind}(R')$ is expressed, considering the polarizability $\alpha_{neutral}$ of the neutral species, adopting the following formalism:
\begin{equation}
   V_{ind}(R') = -7200 \frac{\alpha_{neutral}}{R'^4}
   \label{eq:10}
\end{equation}
By analogy to $R$, $R'$ is the distance between the neutralized species and the center-of-mass of the newly formed cation.
At last, when comparing the ionization energy of the molecular orbitals of the molecule and of the ion, it is possible to estimate from which orbital(s) of the molecule the electron might be removed.
The difference between the two energies must be taken into account when defining the exit PES in order to see if the cuts of the entrance and exit PESs are crossing each other at some point.
This difference in energy is called the asymptote. 
Fig.~\ref{fig_pes} shows the algorithm of the \textit{pes.c} code. In green are represented the input information, such as the spherical coordinates of the ion.
In blue are reported the functions code in the program, such as the equations previously reported.
The outputs, represented in grey, contain the values of the potential for different configurations used to obtain the PESs.
With this program, it is possible to have different cuts of the entrance and exit PESs, and so it is possible to visualise which part of the neutral molecule can be easily approached by the ionic species.

\subsection{Finding the point of crossing with \textit{pesZeri.c}}
\begin{figure}[t]
\centering
\includegraphics[width=\linewidth]{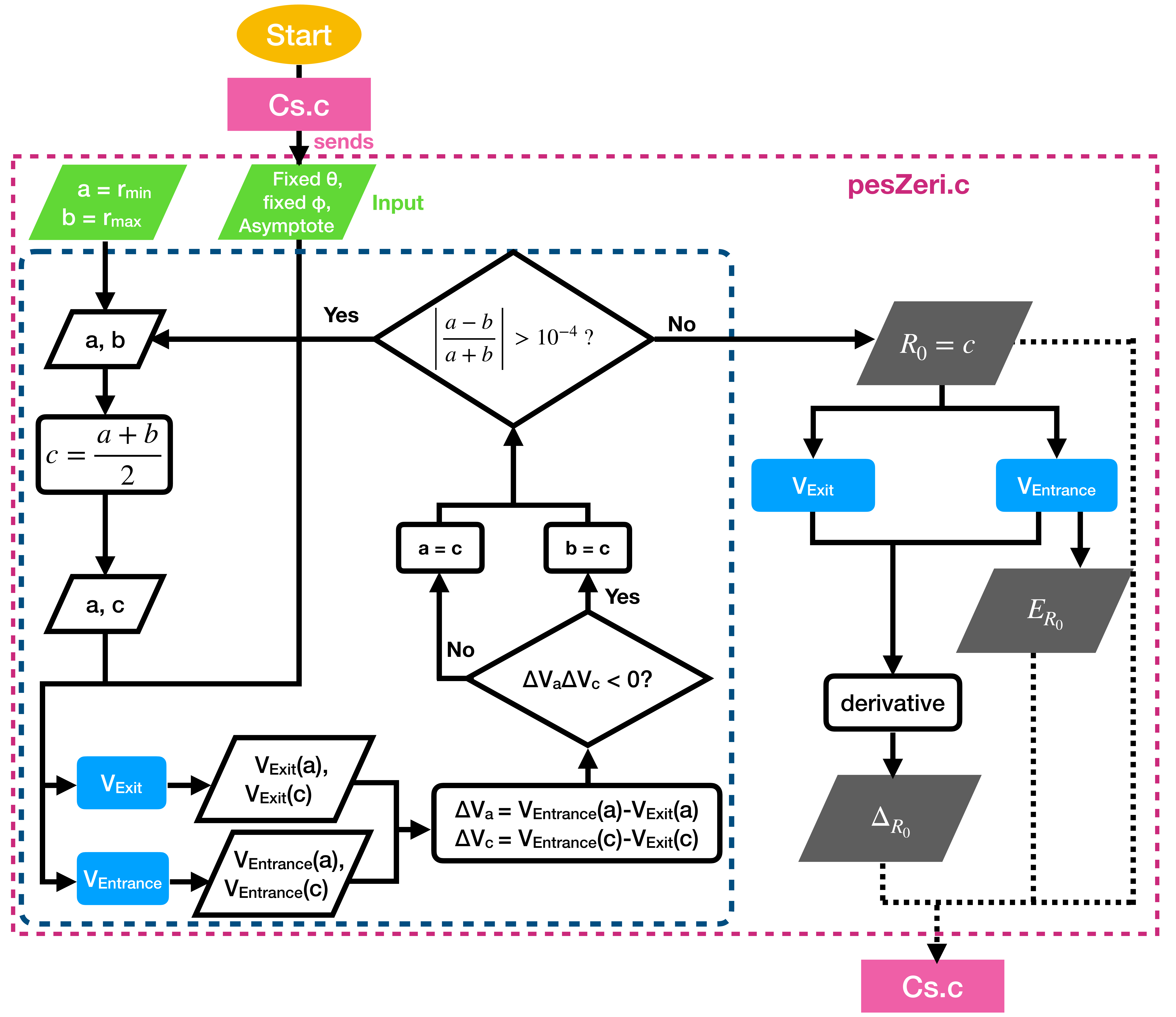}
\caption{Inside the \textit{pesZeri.c} code. The region within the blue dotted frame represents a loop that works until the point of crossing between the entrance and exit PES is found.}
\label {fig_pesZero}
\end{figure}

The piece of code called ``\textit{pesZeri.c}" is basically the same as ``\textit{pes.c}", with the difference that it is used to find the point of crossing between the entrance and exit PES cuts, for a specific configuration ($\theta$ and $\phi$ pair).
As it can be seen in Fig.~\ref{fig_pesZero}, the code enters a loop (space within the dotted blue frame) in order to find the value of $R$ at which the crossing happens ($R_0$).
%
Later, the value of the entrance (or exit) potential at the crossing point, denoted $E_{R_0}$, is also calculated.
Finally, the difference between the slopes of the two PESs cuts is estimated through the formula:
\begin{equation}
   \Delta_{R_0} = \left | \frac{dV_{Exit}(R_0)}{dR}-\frac{dV_{Entrance}(R_0)}{dR} \right |
   \label{eq:11}
\end{equation}

When the calculation ends, the three quantities represented in grey in Fig.~\ref{fig_pesZero} are sent to a code called ``\textit{CS.c}", that will be presented in the next section, and the values will be used to estimate the cross section at different collisional energies.

\subsection{Estimating the transition probability for a configuration using \textit{CS.c}}
\begin{figure}[t]
\centering
\includegraphics[width=0.9\linewidth]{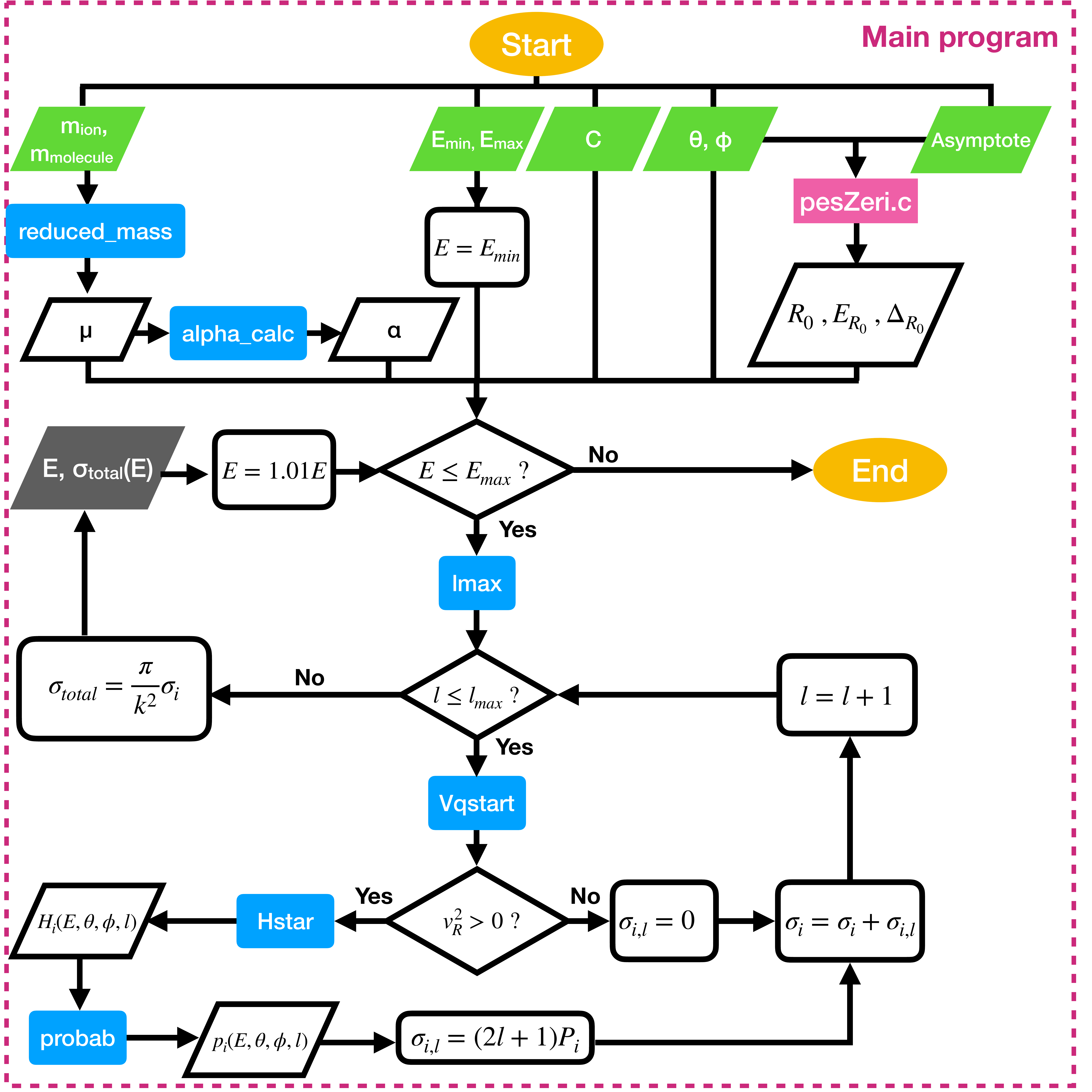}
\caption{Inside the \textit{CS.c} code. }
\label {fig_cs}
\end{figure}

The transition probability at the crossing point between the two PES cuts is treated adopting a a strategy successfully used for previous investigations\cite{candori2001structure,candori2003state,cernuto2017experimental}. In details, the Landau-Zener-Stückelberg approach\cite{Landau1932Theory,zener1932non,Stuckelberg1932theory,Nikitin1984theory,nikitin1999nonadiabatic} is used for the implementation of a one-dimensional model, that is considering specific cuts of the multidimensional PES, where $R$ is the reaction coordinate, keeping the values of $\theta$ and $\phi$ fixed. 

Fig.~\ref{fig_cs} shows the algorithm of the program. At the start, the code reads the input file, which should contain the configuration ($\theta$, $\phi$), the value of the asymptote, the value of $C$ (parameter used in Coriolis coupling that will be shown later), the molecular masses of the reactants and the minimum and maximum collisional energies at which the cross section is estimated.
($\theta$,$\phi$) and the asymptote values are sent to the \textit{pesZeri.c} code, as described before, and in return the program gets the position of the crossing point, the value of the potential at this point and the difference between slopes ($\Delta_{R_0}$ from eq. 11).
Meanwhile, the program calls a function to compute the reduced mass $\mu$, in $kg$, which is then used to calculate $\alpha$.
Here $\alpha$ is a parameter that includes conversion factors: 
\begin{equation}
   \alpha= 10^{-10} \cdot \sqrt{1.62176 \cdot 10^{-22}} \cdot \sqrt{100} \cdot \frac{\sqrt{2 \mu}}{\hbar}  
   \label{eq:122}
\end{equation}
where the first term is the factor of conversion from $m^{-1}$ to \textit{\AA}$^{-1}$, the second term converts $meV^{\frac{1}{2}}$ to $J^{\frac{1}{2}}$, the third term converts ${\frac{meV}{100}}^{\frac{1}{2}}$ to $meV$ and $\hbar$ is the reduced Planck's constant.

The subsequent steps of the program consists in a series of loops. 
In the outer loop, the collisional energy $E$ is set to $E_{min}$, defined by the user in units of $\frac{meV}{100}$.
$E$ will be multiplied by a factor of 1.01 at the end of every iteration, until it reaches the value of $E_{max}$, also defined by the user of the code.
With the collisional energy $E$ and the $\alpha$ parameter, the wavenumber $k$, not show on Fig.~\ref{fig_cs}, is calculated as follows:
\begin{equation}
   k =\alpha \sqrt{E} 
   \label{eq:12}
\end{equation}

The wavenumber $k$ is then used to compute the maximum value of the angular momentum $l_{max}$, for which the system is able to reach the crossing point by overcoming the centrifugal barrier is given by: 
\begin{equation}
   l_{max} = k R_0 \sqrt{1-\frac{E_{R_0}}{E}}
   \label{eq:13}
\end{equation}

With this the program enters a second, inner loop where the value of $l$, the angular momentum quantum number of the collision complex, starts at zero and is increased by one unit until it becomes equal to $l_{max}$.
Whenever $l$ increases, \textit{i.e} in every cycle of the routine, the code estimates 3 quantities:
the radial velocity $v_R$, the non-adiabatic coupling $H_i$ and the probability of crossing $p_i$.
The radial velocity $v_R$ is defined as:
\begin{equation}
   v_{R}^2(l,E) = \frac{2}{\mu}\bigg[E\bigg(1-\frac{l(l+1)}{k^2 R_0^2}\bigg) -E_{R_0} \bigg]
   \label{eq:14}
\end{equation}
%
%
All the quantities appearing in the equation have been defined before.
In the systems the code was written for, the non-adiabatic coupling was treated as a Coriolis coupling, due to the fact that the electron is exchanged between orbitals of different symmetry. The Coriolis coupling expression is
\begin{equation}
   H_i(E,\theta,\phi,l) = \frac{\hbar l}{\mu R^2_0}M
   \label{eq:15}
\end{equation}
The $M$ term has been extended to account for the dependence of the coupling on the strong stereochemistry. In systems involving helium cation and dimethyl ether or methyl formate \cite{cernuto2017experimental,Cernuto2018}, the expression of $M$ takes the form, with a dimensionless parameter $C$:
\begin{equation}
   M = C  {\left| \frac{E}{E_{R_0}} \right |}^{\frac{1}{4}}
   \label{eq:16}
\end{equation}

The probability for the passage through a crossing between entrance and exit PESs is given by:

\begin{equation}
   p_{i}(E,\theta, \phi, l) = exp \bigg(\frac{-2\pi H_i^2}{\hbar v_R  \Delta_{R_0}}\bigg) 
   \label{eq:17}
\end{equation}


Finally, a conditional statement is reached: when the square
radial velocity $v_R^2$ is smaller than or equal to zero, the total cross section $\sigma(E,\theta, \phi)$ is also zero. Otherwise, the cross section is calculated as follows:
\begin{equation}
   \sigma_{total}(E,\theta, \phi) = \frac{\pi}{k^2}\sum_{l=0}^{l_{max}}(2l+1)P_i
   \label{eq:18}
\end{equation}

where $P_i$, the probability of formation of the molecular ion, is expressed taking into account the previously described  $p_i(E,\theta,\phi,l)$ and $l_{max}$ is the maximum value of $l$ for which the system is able to reach the crossing point, through the overcoming of the centrifugal barrier.
In the case of a single crossing between entrance and exit channels $P_i$ assumes the form:
\begin{equation}
   P_i(E,\theta, \phi, l)=(1-p_i)(1+p_i)
   \label{eq:19}
\end{equation}
Once the two loops are completed, the code returns an array containing the cross section for each collisional energy for a specific configuration.
It is also possible to make the code compute the cross section for several configurations and then calculate the average cross section. In this case, the input file must contain all the configurations the user as selected.
In the next section, it is explained how to make an input containing configurations equally distributed in space.

\subsection{Generating an input with \textit{sphere.c} for multiple configurations}
%
%
\begin{figure}[t]
\centering
\includegraphics[width=\linewidth]{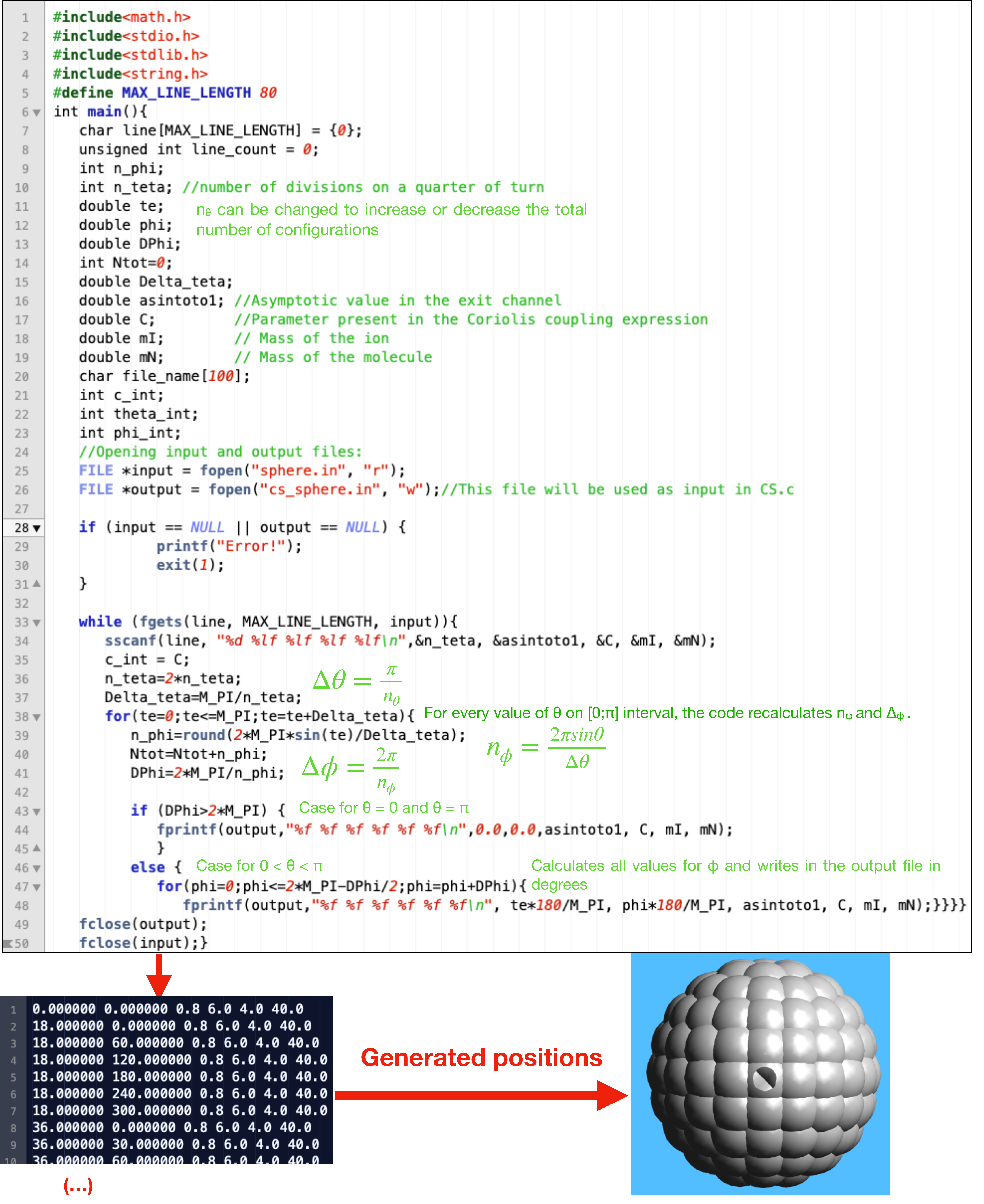}
\caption{Inside the \textit{sphera.c} code}
\label {fig_sphere}
\end{figure}
The purpose of \textit{sphere.c} is to generate an input file for \textit{CS.c}.
The idea is to have several configurations evenly distributed around the neutral species.
For instance, in Fig.~\ref{fig_sphere}, by assuming $n_{\theta}$ = 10, the program returns 128 configurations distributed evenly in a spherical manner.
The input can be used to calculate the average cross section over the three-dimensional space.
This information allows to predict whether the two reactants assume a preferential relative orientation when they are close to each other.

\section{Conclusions}
The pieces of code described have been used in three systems so far, all of them involving He$^+$ colliding with dimethyl ether, methyl formate and methanol.
They have been able to reproduce the data obtained through laboratory measurements of the total cross section as a function of the collision energy for the three systems.
The code, and the physical model at its base, is of relevance in interpreting charge exchange processes at thermal and subthermal collision energies, since it presents several advantages with respect to classical capture models in the calculation cross sections:
\begin{enumerate}
    \item{It includes an estimate of the initial molecular orbital from which the electron is captured by the atom.}
    %
    %
    \item{The detailed exploration of the interaction anisotropy, controlling the relative stability of different sets of configurations, suggests the formation of the collision complex, during the approach of the reagents, in preferential relative orientations, and how such orientations change their weight when the collision energy increases.}
    \item{By using experimental values of the cross sections as benchmark data, the model can estimate the reaction rates as a function of the temperature.}
\end{enumerate}
Eventually, the reaction rate values can be used in astrochemical models in order to predict the chemical evolution of interstellar environments.

\section{Acknowledgements}

    The authors thank Andrea Cernuto who originally developed the code.
    This project has received funding from the European Union’s Horizon 2020 research and innovation programme under the Marie Sk{\l}odowska Curie grant agreement No 811312 for the project ”Astro-Chemical Origins” (ACO).
    The authors thank the Herla Project (http://www.hpc.unipg.it/hosting/vherla/vherla.html) - Universit\`{a} degli Studi di Perugia for allocated computing time.
    The authors thank the Dipartimento di Ingegneria Civile ed Ambientale of the University of Perugia for allocated computing time within the project “Dipartimenti di Eccellenza 2018-2022”.
    N. F.-L thanks MIUR and the University of Perugia for the financial support of the AMIS project through the “Dipartimenti di Eccellenza” programme.
    N.F.-L. also acknowledges the Fondo Ricerca di Base 2021 (RICBASE2021FAGINAS) del Dipartimento di Chimica, Biologia e Biotecnologie della Università di Perugia for financial support.
    D.A. and M.R. acknowledge funding from MUR PRIN 2020 project n. 2020AFB3FX.
    
\clearpage

%
%
%
%

 \bibliography{cs_code}{}
 \bibliographystyle{splncs04}

\end{document}